\documentclass[aps,prl,preprint,superscriptaddress]{revtex4-1}

\usepackage{graphicx}
\usepackage{color}

\newcommand{\be}{\begin{displaymath}}
\newcommand{\bn}{\begin{equation}}
\newcommand{\bea}{\begin{eqnarray*}}
\newcommand{\eea}{\end{eqnarray*}}
\newcommand{\en}{\end{equation}}
\newcommand{\ee}{\end{displaymath}}

\newcommand{\p}{\partial}
\newcommand{\lang}{\left\langle}
\newcommand{\rang}{\right\rangle}

\begin{document}



\title{Impurity transport in a mixed-collisionality stellarator plasma}

\author{P. Helander}
\affiliation{Max-Planck-Institut f\"{u}r Plasmaphysik, 17491 Greifswald, Germany}
\author{S. L. Newton}
\affiliation{Department of Physics, Chalmers University of Technology, SE-412 96 G\"{o}teborg, Sweden}
\affiliation{CCFE, Culham Science Centre, Abingdon, Oxon OX14 3DB, UK}
\author{A. Moll\'{e}n}
\affiliation{Max-Planck-Institut f\"{u}r Plasmaphysik, 17491 Greifswald, Germany}
\author{H.M. Smith}
\affiliation{Max-Planck-Institut f\"{u}r Plasmaphysik, 17491 Greifswald, Germany}

\date{\today}


\begin{abstract}

A potential threat to the performance of magnetically confined fusion plasmas is the problem of impurity accumulation, which causes the concentration of highly charged impurity ions to rise uncontrollably in the center of the plasma and spoil the energy confinement by excessive radiation. It has long been thought that the collisional transport of impurities in stellarators always leads to such accumulation (if the electric field points inwards, which is usually the case), whereas tokamaks, being axisymmetric, can benefit from ``temperature screening'', i.e., an outward flux of impurities driven by the temperature gradient. 
Here it is shown, using analytical techniques supported by results from a new numerical code, that such screening can arise in stellarator plasmas too, and indeed does so in one of the most relevant operating regimes, where the impurities are highly collisional whilst the bulk plasma is in any of the low-collisionality regimes.
\end{abstract}

\pacs{}

\maketitle




\emph{Introduction}. 
Tokamaks and stellarators are the two most developed concepts for magnetic-confinement fusion. They have relative advantages and disadvantages, which have been explored and discussed extensively over the years \cite{helander2012}. For instance, the tokamak requires current drive and suffers from ``disruptive'' instabilities, whereas fast-ion confinement is difficult in stellarators. A further problem for the stellarator is the threat of heavy-ion impurity accumulation in the core of the plasma~\cite{Taylor,Connor,Rutherford,W7-A,Igitkhanov,Hirsch,Ida}.  Maintaining a fusion plasma at the necessary multi-keV temperature requires excellent boundary control, and any potential penetration of impurity ions released in plasma-wall interactions must be dealt with. Unchecked, the radiation from any significant build-up of partially ionized impurities will prevent power balance and quench any fusion reaction. 

The transport of impurity ions in a tokamak or stellarator plasma is governed by turbulent and ``neoclassical'' processes, the latter being caused by the random walk executed by these particles as they travel along complicated orbits set by the magnetic-field geometry whilst colliding with other particles~\cite{Hirshman,helandersigmar2002,helander2014}. The turbulent transport often dominates, but the neoclassical transport can be very significant for heavy impurities in both types of device. Moreover, it is usually in the direction of the bulk-ion density gradient, i.e. {\em inward}, into the core of the plasma. To make things worse, this inward neoclassical transport has been predicted to be particularly strong, practically inevitable, in stellarators. This has been an issue of great concern for several decades. The aim of the present Letter is to show that the situation is less serious than previously thought. 

The neoclassical impurity flux is of the form \cite{Hirshman,Igitkhanov}
\bn
\Gamma_z = \lang \int f_z ({\bf v}_{d} \cdot \nabla r) d^3v \rang
= n_z \left( D_{11}^{zi} A_{1i} + D_{11}^{zz} A_{1z} + D_{12}^{z} A_{2i} \right),
\label{eqdeffluxcoeffs}
\en
where $z$ refers to impurity ions, $i$ to the bulk (hydrogenic) ions, ${\bf v}_{d}$ denotes the drift velocity, $r$ is an arbitrary label of the magnetic surfaces serving as radial coordinate, angular brackets indicate an average over such surfaces, and
$$
A_{1a} = \frac{d \ln p_a}{d r} + \frac{e_a \phi'(r)}{T_a}, \qquad A_{2a} = \frac{d \ln T_a}{d r}
$$
denote the ``thermodynamic forces''. Here $p_a = n_a T_a$ is the pressure of species $a$, $e_a$ its charge, and $\phi(r)$ the electrostatic potential, which like the density $n_a$ and temperature $T_a$ is approximately constant on magnetic surfaces. Different ion species usually have the same temperature, so for simplicity we have $A_{2i} = A_{2z}$. In axisymmetric magnetic fields, the largest transport coefficients are $D_{11}^{zi}$ and $D_{12}^{z}$, where the former is always positive and so tends to drive impurities into the plasma if $d p_i/dr < 0$. The coefficient $D_{12}^{z}$ is usually negative and can (depending on the collision frequency) exceed $D_{11}^{zi}$ \cite{Connor}. If the ion temperature profile is sufficiently steep, outward impurity transport will then result. This beneficial property is referred to as {\em temperature screening}. In tokamaks, the transport coefficients also have the property that the sum of all terms containing the radial electric field, ${\bf E} = - \phi'(r) \nabla r$ vanish. 

However, for heavy impurities ($a = z$, $e_a = Ze$, $m_z/m_i = O(Z) \gg 1$) in a stellarator the picture has been pessimistic, because the radial electric field term {\em does} contribute to the transport. Since this term is multiplied by the large number $Z \gg 1$ in $A_{1z}$, it tends to dominate and will drive the impurities in the direction of the electric field, which typically points inward in fusion-relevant high-density plasmas. Unless there is very strong turbulence, which impairs energy confinement, any high-$Z$ impurity will thus accumulate in the core, as is indeed often observed in experiments~\cite{W7-A,Igitkhanov,Hirsch}. A notable exception occurs in low-density ``impurity-hole'' plasmas in the Large Helical Device \cite{Ida}.

These conclusions were mostly based on calculations that approximate the collisions between different particles species by a simple scattering operator, sometimes augmented by a term ensuring momentum conservation. This treatment is adequate when all ions are at low collisionality (set by the ratio of the major radius to the mean free path and denoted by $\nu_*^{ab}$ for collisions between two species $a$ and $b$). However, the regime of greatest practical interest is the one where the mean free path of the bulk ions is long but that of high-$Z$ impurities relatively short. (It scales as $Z^{-4}$ at constant impurity density.) This mixed-collisionality regime cannot be treated correctly by numerical codes that neglect interaction other than collisional scattering between different ion species. For several decades, such codes have been the work horses for neoclassical transport calculations in stellarators~\cite{beidleretal2011}. 

We therefore consider the neoclassical transport in a mixed-collisionality plasma with a single, highly charged impurity species, and calculate the cross-field flux $\Gamma_z$ by solving the kinetic equation for the distribution function $f_z$ both analytically and numerically, using the full Landau collision operator for the impurities. The numerical calculation can be done for arbitrary collisionality, but the analytical treatment is only possible by considering the asymptotic limit of short and long mean free paths, respectively, for the impurities and bulk ions. More details of both calculations will be published separately.


\emph{Analytical calculation}. 
In order to evaluate the impurity particle flux analytically, it is useful to decompose it into a sum of contributions~\cite{sugamanishimura2002,braunhelander2010}, driven by the friction against the background bulk ions and the pressure anisotropy,
	\bn \Gamma_z = \frac{1}{Ze} \lang u B R_{zi\parallel} + (p_{z\|} - p_{z\perp} ) \frac{\nabla_\| (uB^2)}{2 B} \rang. 
	\label{flux-friction relation}
	\en
Here, the second term on the right is relatively small 
 for a highly collisional population, with $T_z = T_i \equiv T$, and can be neglected when the collisionalities satisfy $\nu_*^{iz} \nu_*^{zz} \gg n_z  Z^{1/2} / n_i$, 
a condition we take to hold in the analytic calculation. The remaining term contains the function $u$, which is related to the parallel (to the magnetic field) plasma current and pressure by $u = J_\|/ (B p'(r))$, and the friction force 
		\bn R_{zi\parallel} = \int m_z v_\parallel C_{zi}(f_z,f_i) d^3v = m_i \int \nu_D^{iz}(v) v_\parallel f_i d^3v - \frac{m_in_i V_{z\parallel}}{\tau_{iz}}, 
		\label{friction}
		\en
where the deflection frequency is $\nu_D^{iz}(v) = (3 \pi^{1/2} / 4\tau_{iz}) (v_{Ti}/v)^3$ and the collision time $\tau_{iz} = 3(2\pi)^{3/2} \sqrt{m_i} T^{3/2} \epsilon_0^2/(n_z Z^2 e^4 \ln \Lambda)$ \cite{helandersigmar2002}. To evaluate the first term appearing in Eq.~(\ref{friction}), we require the solution of the kinetic equation for the distribution function $f_i$, specifically the piece which is odd in the parallel velocity, $v_\parallel = \sigma |v_\parallel|$.

A recently developed formulation allows a unified treatment of this problem throughout the low-collisionality regimes of the bulk ions~\cite{helanderetalsub}. If  the distribution function is split into even and odd pieces, $f^\pm$, the kinetic equation becomes $v_\parallel \nabla_\parallel f^{\mp} = C^{\pm}(f) - {\bf v}_d \cdot \nabla f^{\pm}$, where the independent coordinates are $\left(r, \alpha, l, \epsilon, \mu, \sigma \right)$. Here $\alpha$ is a label for the different field lines on the same flux surface, $l$ the arc length along the magnetic field, $\epsilon = m_i v^2 / 2 + e_i \phi({\bf x})$ the energy, and $\mu = m_i v_\perp^2/2B$ the magnetic moment. The electrostatic potential can be set to $\phi = 0$ on the surface of interest, and particle orbits are defined as passing (able to move over a whole flux surface) or magnetically trapped depending on whether the parameter $\lambda = \mu / \epsilon$ is less or greater than $1/B_{\rm max}(r)$, where $B_{\rm max}(r)$ is the maximum field strength on the surface. The odd part of the bulk-ion distribution function can thus be written as
\bn
f^- \left(r, \alpha, l, \epsilon, \mu, \sigma \right) = \int_{l_0}^l \left[C^+(f) - {\bf v}_d \cdot \nabla f^+\right] \frac{dl^\prime}{v_\parallel} + X,
\label{eqformfodd}
\en
where $X\left(r,\alpha,\epsilon,\mu,\sigma\right)$ denotes an integration constant that is independent of $\alpha$ in the passing region and vanishes in the trapped region if $l_0$ is chosen to be a bounce point. At such points $\lambda = 1/B$, the parallel velocity vanishes, and so therefore must $f^-$. 

The parallel streaming term in the kinetic equation is annihilated by the orbit average, which is defined as a time average taken along a trajectory following the magnetic field between two consecutive bounce points for trapped particles, or many times around the torus for passing ones. (Formally, this average is obtained by multiplying the kinetic equation by $dl/v_\|$ and integrating.) This gives the equation 
	$$\overline{{\bf v}_d \cdot \nabla f^+} = \overline{C^+(f)} $$ 
for $f^+$, where the orbit average is indicated by $\overline{(\cdots)}$. For passing particles and for trapped ones at moderately low bulk-ion collisionality (the $1/\nu$-regime), the right-hand side dominates, and the solution is approximately Maxwellian,  $f^+ \simeq F_0(r,\epsilon)$. At lower bulk-ion collisionality (the $\sqrt{\nu}$-regime), trapped-particle drifts produce strong deviations from such an equilibrium and prevent effective plasma confinement unless the magnetic field is optimized to be nearly omnigeneous~\cite{calvoetal2016}, or the radial electric field is sufficiently strong to produce an in-surface ${\bf E} \times {\bf B}$ drift which averages out the magnetic drift motion over an orbit~\cite{hokulsrud1987}. In both cases, particles stay close to a flux surface on an orbit average, so that $f^+ = F_0 + F_1$ is determined by 
	$$\overline{{\bf v}_d \cdot \nabla \alpha} \, \frac{\p F_1}{\p \alpha} + \overline{{\bf v}_d \cdot \nabla r} \, \frac{\p F_0}{\p r} \simeq 0. $$ 
This equation holds for all trapped particles except those in a thin layer around the trapped-passing boundary, which is unimportant for our present purposes but regulates the transport of the bulk ions \cite{hokulsrud1987}. We thus conclude that ${\bf v}_d \cdot \nabla f^+ \simeq  ({\bf v}_d \cdot \nabla r) \p F_0 / \p r$ in the $1/\nu$-regime and ${\bf v}_d \cdot \nabla f^+\simeq ({\bf v}_d \cdot \nabla r- \overline{{\bf v}_d \cdot \nabla r}) \p F_0 / \p r$ in the $\sqrt{\nu}$-regime \cite{helanderetalsub}. 

We now evaluate the friction force between the main ions and the impurities by substituting the solution (\ref{eqformfodd}) in Eq.~(\ref{friction}). The collision term in Eq.~(\ref{eqformfodd}) is small in the $\sqrt{\nu}$-regime but generally produces a contribution to the friction in the $1/\nu$-regime. However, a commonly used, and successful~\cite{landremanetal2014}, approximation to the collision operator for transport applications takes the form of a pitch-angle scattering operator with a momentum-conserving term, in which case this contribution vanishes. We will therefore neglect it, but its impact can be tested in the final numerical comparison. The remaining terms give 
	$$ R_{zi\parallel} = \frac{p_i m_i}{e \tau_{iz}} \left(A_{1i} - \frac{3}{2}A_{2i}\right) (u+s)B + P(r) B - \frac{m_in_iV_{z\parallel}}{\tau_{iz}}, $$
where terms that vary as $B$ over the flux surface have been combined into $P(r)$, a flux surface function that turns out not to affect the impurity flux. In deriving this result, we have written ${\bf v}_d \cdot \nabla r = \Omega_i^{-1} v_\parallel \left( {\bf B} \times \nabla r \right)\cdot \nabla \left(v_\parallel / B\right)$, and defined $s=0$ in the $1/\nu$-regime and
\bn
s(l) = \frac{3}{2} \int_{l_{\rm max}}^l dl^\prime
\int_{1/B_{max}}^{1/B(l')} \frac{d\lambda}{\xi(l')} \; \overline{\xi \left(\hat{\bf b} \times \nabla r \right) \cdot \nabla \left(\frac{\xi}{B}\right)} 
\en
 in the $\sqrt{\nu}$-regime, with $\xi = \sqrt{1-\lambda B}$.

The impurity flow $V_{z\parallel}$ takes the well-known form $n_z V_{z\parallel} = (p_z/Z e)A_{1z} uB + K_z(r) B$, where $K_z$ is an integration constant determined by force balance along the magnetic field, which reduces to $\lang BR_{zi\parallel} \rang = 0$ in the collisional limit~\cite{hirshmansigmar1977}. 
Using this relation to determine $K_z$, we eliminate the flux function $P(r)$ from the required flux surface average $\lang uBR_{zi\parallel} \rang$ and obtain the final expression for the impurity flux (\ref{flux-friction relation}),
\begin{eqnarray}
&& \Gamma_z = - \frac{m_i p_i}{Ze^2 \tau_{iz}} \left[ \frac{A_{1z}}{Z}\left(\lang u^2 B^2 \rang  - \frac{\lang uB^2\rang^2}{\lang B^2 \rang}\right) \right. 
\nonumber  \\
& & - \left. \left(A_{1i} - \frac{3}{2}A_{2i}\right)\left(\lang u\left(u+s\right)B^2 \rang - \lang \left(u+s\right)B^2 \rang \frac{\lang uB^2 \rang}{\lang B^2 \rang}\right) \right],
\label{eqflux}
\end{eqnarray}
where we recognize the Pfirsch-Schl\"uter diffusion coefficient
	$$ D_{\rm PS} = \frac{m_i T_i}{e^2 \tau_{iz}} \left[\lang u^2 B^2 \rang  - \frac{\lang uB^2\rang^2}{\lang B^2 \rang}\right]
	= \frac{\rho_i^2}{\tau_{iz}} \frac{\lang J_\|^2 \rang \lang B^2 \rang - \lang J_\| B \rang^2}{(dp/dr)^2}, $$
with $\rho_i^2 = m_i T_i / e^2 \lang B^2 \rang$. 

The impurity transport coefficients $D_{ij}^{za}$ appearing in Eq.~(\ref{eqdeffluxcoeffs}) can now be identified, and we note the following points.
The Schwartz inequality implies $D_{\rm PS} \geq 0$, so $D_{11}^{zz}$ is negative, as required by entropy considerations. 
In the $1/\nu$-regime (moderate bulk ion collisionality) $s=0$, and we see $D_{11}^{zi} = -Z D_{11}^{zz}$, as is also true in the very-high-collisionality limit where both ion species are collisional \cite{braunhelander2010}. The direct drive of the flux by the electric field thus {\em cancels out} in both of these regimes. Furthermore, we note that $D_{12}^{z} = -(3/2)D_{11}^{zi}$, so there is {\em temperature screening} in the $1/\nu$-regime. 
At lower bulk ion collisionality, the electric-field drive no longer cancels exactly, becoming proportional to the quantity $\lang usB^2\rang - \lang sB^2\rang\lang uB^2\rang / \lang B^2\rang$, which must be evaluated numerically. 
The same additional contribution, due to the trapped particle drift, appears in the transport coefficient multiplying $dT/dr$, but the relation $D_{12}^{z} = -(3/2)D_{11}^{zi}$ continues to hold, so there is either temperature screening or outward transport due to the bulk-ion density gradient, depending on the sign of the last term in Eq.~(\ref{eqflux}). 

This analytical calculation thus shows that (i) there can be temperature screening both in the $1/\nu$- and $\sqrt{\nu}$-regimes and (ii) the radial electric field may only weakly drive impurity transport. Both results are at odds with conventional wisdom. 

\emph{Numerical calculation.}
A novel computational tool has recently been developed, the continuum $\delta f$ SFINCS code, which solves the coupled first-order drift-kinetic equations in general magnetic geometries and calculates the neoclassical transport for an arbitrary number of species, retaining the full linearized multi-species Landau collision operator. The numerical implementation, which includes the calculation of perturbed Rosenbluth potentials, is detailed in~\cite{landremanetal2014}.
SFINCS was used by Moll\'{e}n {\it et.~al.}~\cite{mollenetal2015} in an extensive study of the transport of impurities and their effect on the bootstrap current in the recently completed Wendelstein 7-X stellarator \cite{klingeretal2017}. Puzzling indications of temperature screening were seen already in this work, which we are now able to understand and clarify in terms of the analytical theory above and further numerical results.

Figure 1 shows an example where the transport coefficients for commonly occurring C$^{6+}$ impurity ions in a bulk H$^+$ plasma were studied over a wide range of impurity collisionality, defined as $\nu_\ast^{zz} = R / (v_{Tz} \tau_{zz})$ with $\tau_{zz} = (m_z/m_i)^{1/2} Z^{-2} \tau_{iz}$ and $R = (G + \iota I)/B_{00}$, where the magnetic field is ${\bf B} = K \nabla \psi + I \nabla \theta + G \nabla \varphi$ in Boozer coordinates and $B_{00}$ denotes the $m=n=0$ Fourier harmonic of the field strength $B(\psi,\theta,\varphi)$. In both Figs. 1a) and 1b), $E_r = 0$ and, as expected from the analytical calculation, the coefficient $D_{11}^{zz} + D_{11}^{zi} + D_{12}^z$ multiplying the temperature gradient is negative, implying temperature screening up to the collisionality where the bulk ions leave the long-mean-free-path regime and become collisional, at which point temperature screening disappears \cite{braunhelander2010}. If the collision operator is replaced by pure pitch-angle scattering, however, the temperature screening disappears from the entire range of high impurity collisionality. The range of collisional impurities ($\nu_*^{zz} >1$) and collisionless ($\nu_*^{iz} <1$) bulk ions is demarked by the thick black line, which gives the value of the coefficient obtained from Eq.~(\ref{eqflux}). Note that this value matches the results from both SFINCS (with the full Landau collision operator) and the DKES code if a momentum-conserving term is added to the scattering operator~\cite{maassberg}. However, the latter treatment fails in the very-high-collisionality limit where all species are collisional. This limit can be treated analytically \cite{braunhelander2010} and the asymptote is shown as a dashed red line, along with that for pitch-angle scattering without momentum correction, matching the corresponding numerical results in the appropriate limit.

The coefficient $D_{11}^{zz} + D_{11}^{zi} + D_{12}^z$ is negative also at low impurity collisionalities, but its contribution to the transport is expected to be overwhelmed by the inward flux produced by a radial electric field. The corresponding electric-field transport coefficient (the field is assumed to be weak enough to allow the $1/\nu$-regime over the collisionality range) is shown in Fig.~1b) and can be seen to be large at low collisionality but small in the mixed collisionality regime, enabling temperature screening. 

\begin{figure}
\includegraphics[width=\textwidth]{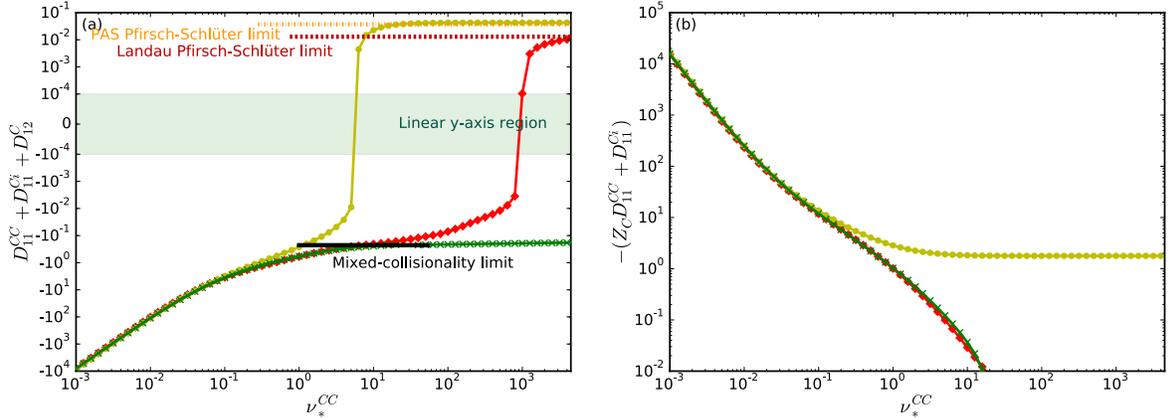}
\caption{Normalized transport coefficients of C$^{6+}$ impurity flux driven by a) the bulk ion temperature gradient, $D_{11}^{zi} + D_{11}^{zz} + D_{12}^{z}$, and b) the radial electric field, $ZD_{11}^{zz} + D_{11}^{zi}$, in an H$^+$ plasma for $E_r \approx 0$, as a function of impurity self-collisionality at fixed density ratio, in the W7-X standard configuration. The transport coefficients are normalized to 
$n_i \rho_i^2 / (Z n_z \tau_{iz})$, the effective ion charge is $Z_{\rm eff} = \sum_{i,z} n_a Z_a^2/\sum_{i,z} n_a Z_a = 2$, and the normalized radial position $r/a = 0.88$. Red $\diamond$: SFINCS output with full Landau collision operator, yellow $\circ$: SFINCS output retaining only pitch angle scattering in collisions, green $\times$: output from DKES with momentum conservation. Thick black line: value from Eq.~(\ref{eqflux}) for $E_r = 0$, extending over the mixed-collisionality range. Dashed lines: high-collisionality asymptotes from analytical theory using the linearized Landau collision operator and the pitch-angle scattering operator.}
\end{figure}

Figure 2 shows transport coefficients similar to Fig.~1 but for the heavy trace impurity Fe$^{16+}$ and $Z_{\rm eff} = 1.07$. Again, the analytical prediction (\ref{eqflux}) is confirmed, but now there is temperature screening also in the very-high-collisionality regime where both ion species are collisional, as expected for very clean plasmas \cite{Rutherford}. 

\begin{figure}
\includegraphics[width=\textwidth]{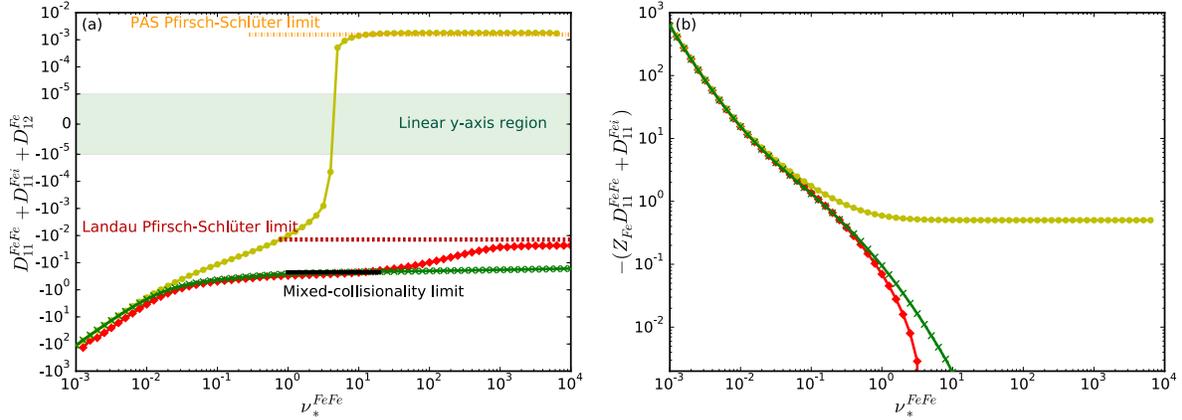}
\caption{Same as Fig.~1 but for Fe$^{16+}$ and $Z_{\rm eff} = 1.07$. There is now temperature screening both in the mixed-collisionality regime and at higher collisionalities, where both ion species are in the Pfirsch-Schl\"uter regime.}
\end{figure}

When a larger radial electric field is present, the terms in Eq.~(\ref{eqflux}) involving the quantity $s$ are important and can be expected to reduce the transport driven by the bulk-ion density and temperature gradients \cite{helanderetalsub}. Temperature screening is still possible, but the contributions from the radial electric field in these terms no longer exactly cancel that from the impurity thermodynamic force $A_{1z}$. On the one hand, there is thus a reduction of the overall neoclassical transport (making it easier, for instance, for turbulence to expel the impurities), but on the other hand some direct electric-field-driven transport now remains. The net transport thus depends sensitively on the collisionality, charge number and the relative size of the density and temperature gradients. By way of example, Fig.~3 shows the transport coefficients in the same magnetic W7-X configuration as for Figs.~1 and 2, but in the presence of a finite radial electric field of $E_r = -5 \rm kV/m$. Temperature screening still occurs (if the temperature gradient is strong enough), but at a lower level, as one would expect from the theory above, making the coefficient comparable to the electric-field coefficient at high collisionality.

\begin{figure}
\centerline{
\begin{minipage}{0.5\textwidth}
\includegraphics[width=\textwidth]{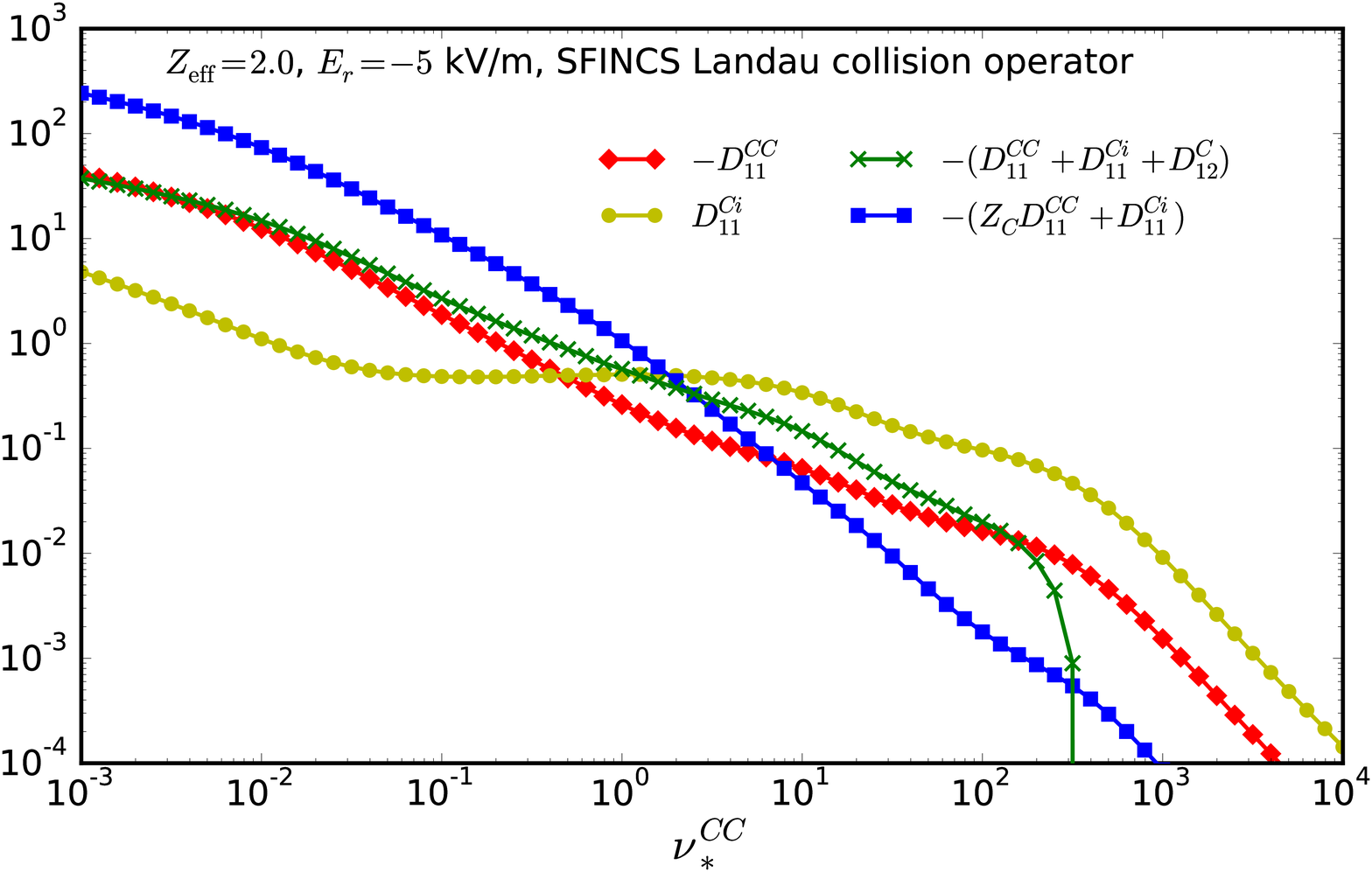}
\end{minipage}\hfill
\begin{minipage}{0.5\textwidth}
\includegraphics[width=\textwidth]{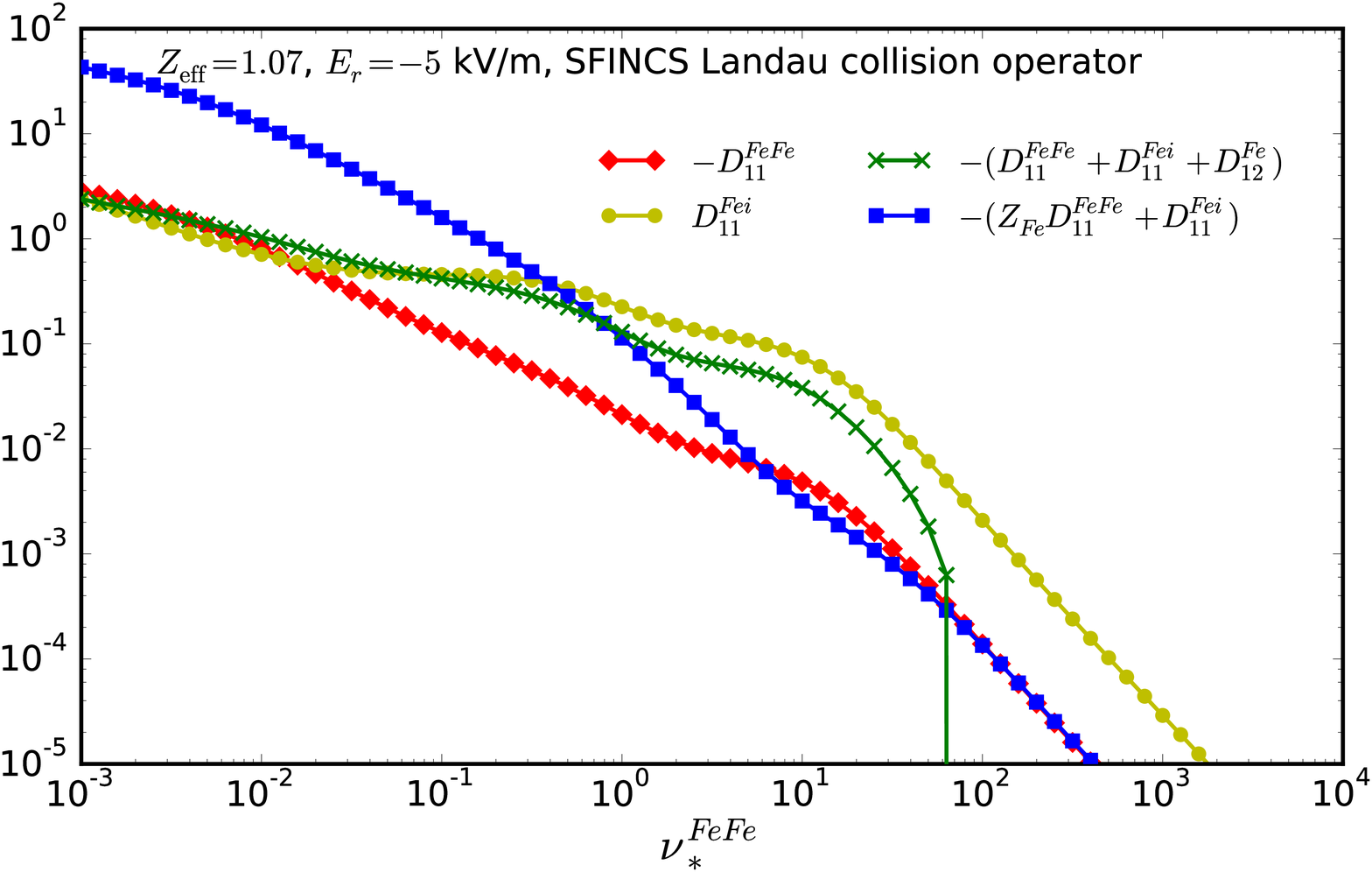}
\end{minipage}
}
\caption{Normalized transport coefficients of C$^{6+}$ and Fe$^{16+}$ impurity fluxes driven by various gradients in the same W7-X standard configuration as in Figs.~1 and 2, but with a radial electric field $E_r = -5$kV/m.}
\end{figure}


\emph{Conclusions.}
In summary, we have found that highly charged, collisional impurity ions in stellarators can experience neoclassical temperature screening if the mean free path of the bulk ions is long and the temperature profile is sufficiently steep, just as in a tokamak. Under these conditions, which are common in experiments, the radial electric field must compete to drive the impurities inward. Impurity accumulation is therefore not inevitable, and impurities may be expected to enjoy outward collisional transport when the conditions are right.



%





\begin{acknowledgments}
We thank Craig Beidler, Felix Parra, Matt Landreman and T\"{u}nde F\"{u}l\"{o}p for helpful discussions, and acknowledge the hospitality of Merton College, Oxford, where this work was initiated.
This work was supported by the Framework grant for Strategic Energy
Research (Dnr. 2014-5392) from Vetenskapsr{\aa}det.
\end{acknowledgments}



\end{document}